\documentclass[epj]{svjour}
\usepackage{graphics}
\usepackage{float}
\usepackage{subfigure}
\usepackage{color}
\usepackage{geometry}
\usepackage{subfig}
\usepackage{caption}
\usepackage{amsfonts}
\usepackage{amsmath}
\usepackage{extarrows}
\usepackage{amssymb}
\usepackage{verbatim}
\usepackage{todonotes}
\usepackage{tikz}
\usepackage{slashed}
\usepackage{mathrsfs}
\usepackage{booktabs}
\usepackage{verbatim}
\usepackage{epsfig}
\usepackage{footnote}
\begin{document}
\title{nonlocal quark condensate from Dyson-Schwinger Equation and its contributions to the gluon vacuum polarization based on OPE approach}
\author{Jing-Hui Huang\inst{1} \and Xue-Ying Duan\inst{2,4,5} \and Chen Huan\inst{3} \and Guang-Jun Wang\inst{2,4,5} \and Xiang-Yun Hu\inst{1*}
}                     
\offprints{}          
\institute{School of Institute of Geophysics and Geomatics, China University of Geosciences, Lumo Road 388, 430074 Wuhan, China. \and School of Automation, China University of Geosciences, Lumo Road 388, 430074 Wuhan, China. \and School of Mathematics and Physics, China University of Geosiciences, Lumo Road 388, 430074 Wuhan, China. \and Hubei Key Laboratory of Advanced Control and Intelligent Automation for Complex Systems, China \and Engineering Research Center of Intelligent Technology for Geo-Exploration, Ministry of Education, China }

 \mail{$^{1}$xyhu@cug.edu.cn}
\date{Received: date / Revised version: date}
\abstract{
The operator-product expansion(OPE) could be employed to obtain the lowest-order, nonlocal quark scalar condensate component of gluon vacuum polarization. In particular, nonlocal quark scalar condensate can be calculated by solving Dyson-Schwinger Equation(DSE) of QCD. Then, field-theoretic aspects of the gluon vacuum polarization and nonperturbative gluon propagator will be considered in the Landau gauge of the Lorentz gauge fixing. The gluon propagator we obtained is finite in the infrared domain where the single gluon mass $m_g$ can be determined. Our results of the ratio $m_{g}/\Lambda_{QCD}$ the range of that from 1.33 to 1.39 agree with previous determinations for this ratio. Besides, the analytic structure of the gluon propagators from the OPE's result is explored. Our numerical analysis of the gluon' Schwinger function finds clear evidence of the positivity violations in the gluon propagator. In addition, a new method for obtaining the chemical potential dependence of the gluon vacuum polarization and the dressed gluon propagator is developed.
\PACS{
      {PACS-key}{discribing text of that key}   \and
      {PACS-key}{discribing text of that key}
     } 
}
\maketitle


\section{Introduction}
\label{intro}

The Operator Product Expansion (OPE) provides a general framework
to study the nonperturbative regime of Quantum chromodynamics(QCD). OPE introduces the gauge-invariant  condensates into Green functions of QCD \cite{Wilson1969Non,T1989Quark,2005Schwinger,2001Vacuum,Kadavy:2020lng}.
Then, the use of OPE has achieved success in a mass of theoretical analyses in quantum field theories, such as, the Operator Product Expansion dominated by the dimension-two $\langle A^2 \rangle$ condensate is used to fit the running of the coupling\cite{Boucaud:2013jwa}. The gluon propagator model \cite{Shi:2016koj,2011The} incorporating quark’s feedback through operator product expansion (OPE) can be introduced to investigate the QCD phase diagram. Furthermore, the high-energy complicated behaviors of three point Green functions\cite{Kadavy:2020lng} of the QCD chiral currents and densities were studied by the framework of the
operator product expansion in the chiral limit...

Dyson-Schwinger Equation(DSE) approach of QCD~\cite{Beringer:1900zz,Roberts:2000aa,Maris2003,Eichmann:2016yit,Fischer:2018sdj} and Lattice QCD simulation\cite{Bogolubsky:2009dc,Deur:2016tte,Faber:2017alm,Komijani:2020kst} are non-perturbation approaches that describe QCD's several important features, e.g.,the QCD phase diagram, the properties of hadron. But Lattice QCD fails due to the fermion problems when extended to large chemical potential region\cite{Kaczmarek:2011zz,deForcrand:2010he}. Thus, Dyson-Schwinger Equation(DSE) approach of QCD has been effective theory to investigate the finite-density region of QCD matter\cite{YuanChen2006,Chen2008,Chen2009,Huang:2019crt}. 

Recently, considerable progresses~\cite{Fischer2009,Fischer2011,Muller2013pya,Papavassiliou2014} were made in solving the DSE by extracting some of the relevant input from the lattice. Specifically, the quenched or unquenched gluon propagator and a phenomenological ansatz for quark-gluon vertex from Lattice QCD's simulation have successfully been used as input for the DSE calculation. In these researches, the gluon DSE in Fig.(\ref{Fig_GluonDSE}) was solved by assuming that the gluon's vacuum polarization only consists of the contribution of quark loop. In the previous works\cite{Fischer:2012vc,Fischer2009,Fischer2011} show the gluon vacuum polarization at finite chemical potential is the appearance of quadratic divergencies. The technical projection methods, such as the Brown-Pennington projection method\cite{Brown:1988bm}, are quite complex and dependence of a sharp momentum cut-off $\Lambda$ in the quark loop. In this case, the following gluon propagator model\cite{Shi:2016koj,T1989Quark,Jiang:2011up} incorporating quark's feedback through OPE was introduced to investigate the gluon vacuum polarization:
\begin{equation}
\label{old_OPE}
\Pi_{\mu \nu}(p)=(\delta _{\mu \nu}-\frac{p^{\mu}p^{\nu}}{p^2})\frac{g^{2}m_{q}\langle \bar{\psi} \psi \rangle }{3p^{2}}+... \quad q^{2} \gg m_{q}^{2}
\end{equation}

\begin{figure}[t]
\vspace*{-5mm}
\centering \includegraphics[width=0.35\textwidth]{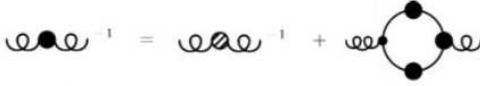}
\vspace*{-5mm}
\caption{Unquenching of gluon propagator. The shaded propagator is the quenched propagator form Lattice QCD.}
\label{Fig_GluonDSE}
\end{figure}
where $m_q$ presents u/d quark, $\langle \bar{\psi} \psi \rangle$ is the gauge-invariant, dimensional objects associated with the Green functions of QCD, and the ellipsis represents terms of higher orders in $m_{q}^{2} / p^{2} $ which are neglected. The strong coupling constant $g^{2}$ is equal to $g^{2}=4 \pi \alpha(\nu)$  with $\alpha(\nu)$ = 0.3 representing the running coupling at a scale fixed by the quenched gluon from the lattice\cite{2004A}. It is worth noting that, the model(Eq.\ref{old_OPE}) shows the value of gluon vacuum polarization varies widely due to the uncertainty of quark current mass $m_q$ and the asymptotic form(Eq.\ref{old_OPE}) makes it inappropriate to study the behaviour of gluon vacuum polarization in the deep infrared domain.

One of the most intricate problems in quantum field theories is the separation of physical and nonphysical degrees of freedom. In vacuum or at finite chemical potential, the ground state in the DSE frame of QCD is characterized  with  confinement  and  dynamical  chiral  symmetry breaking (DCSB), i.e. The analytic structure of the Schwinger function from numerical solutions of DSEs in the space-like Euclidean momentum region have been investigated by searching for the clear evidence of  positivity violations \cite{vonSmekal:1997ohs,Fischer:2003rp,Alkofer:2003jj} in the gluon and quark propagator.

Inspired by the success of Dyson-Schwinger Equation and OPE approach, we further study in this paper the gluon vacuum polarization expanded to the nonlocal quark condensate rather than the local one. The nonlocal quark condensate can be calculated by solving the quark's DSE\cite{Chen2008,Huang:2019crt,Roberts:2000aa}. Then, the unquenched gluon propagator is obtained and compares to the lattice result\cite{Bowman:2007du,Ayala:2012pb}. Furthermore, we expand these results to the case at finite chemical potential. On the other hand, the Green functions component of the gluon vacuum polarization is an important field-theoretic test of DSE approach and OPE techniques, since our unquenched gluon propagator is consistent with the DSE and can  compare to the Lattice calculations. In addition, the gluon mass and the gluon's Schwinger function will be calculated and compared with previous works\cite{GayDucati:1993fn,Mihara:2000wf,2004A,Bowman:2007du,Ayala:2012pb}.                               
The paper is organized as follows. In Sec.~\ref{SecII}, the truncation schemes of the DSE for the quark propagator in vacuum and at finite chemical potential are given. In Sec.~\ref{SecIII}, the nonlocal quark condensate component of gluon vacuum polarization and the unquenched gluon propagator are evaluated.Then, the numerical results are given in Sec.~\ref{SecIV}. Finally, we summarize our work and conclude with a brief remark in Sec.~\ref{SecV}.

\section{Dyson-Schwinger Equation for the quark propagator }
\label{SecII}
To study the nonlocal quark condensate and the gluon vacuum polarization, we employ the Dyson-Schwinger equation, in which the quark gap equation at finite chemical potential can be written as
\begin{eqnarray}
S(p ; \mu)^{-1}&= &Z_2 (i\gamma\cdot \tilde{p}+m_{q}) 
 +  Z_1 g^2(\mu)\int \frac{d^4q}{(2\pi)^4}
 \nonumber \\ & &\times\!D_{\rho\sigma}(k;\mu) \frac{\lambda^a}{2}\gamma_\rho S(q;\mu)
\Gamma^a_\sigma(q,p;\mu) \, ,
\label{gendse}
\end{eqnarray}
where $\tilde p=(\vec p,p_4+i\mu)$, $k=p-q$, $D_{\rho\sigma}(k ;\mu)$ is the full gluon propagator, $\Gamma^a_\sigma(q,p;\mu)$ is the dressed quark-gluon vertex, $Z_{1}$ is the renormalization constant for the quark-gluon vertex, and $Z_{2} $ is the quark wave-function normalization constant. The general structure of the quark propagator at finite chemical potential can be written as
\begin{eqnarray}
S^{-1}(p;\mu)&=&i\vec{\gamma}\cdot
\vec{p}A(\vec{p}^2,p_4;\mu)+i\gamma_4\tilde{p_4}C(\vec{p}^2,p_4;\mu)
\nonumber \\ & & + B(\vec{p}^2,p_4;\mu) \, ,
\end{eqnarray}
where $A(\vec{p}^2,p_4;\mu)$, $B(\vec{p}^2,p_4;\mu)$,$C(\vec{p}^2,p_4;\mu)$ are scalar functions of $p^2$ and $p_4$.

We solve Eq.~(\ref{gendse}) with models of the gluon propagator and the quark-gluon vertex, which describe meson properties in vacuum well in the symmetry-preserving Dyson-Schwinger equation and Bethe-Salpeter equation (BSE) schemes (see, e.g., Refs.~\cite{Chang2009,BSE,Fischer2009,Binosi:2014aea}).
In vacuum they are usually taken as
\begin{equation}
 { Z_{1}g^2 D_{\rho \sigma}(k) \Gamma_\sigma^a(q,p)}  =
{\cal G}(k^2)D^0_{\rho
\sigma}(k)\frac{\lambda^a}{2}\Gamma_{\sigma}(p,q) \, ,
\label{KernelAnsatz}
\end{equation}
where $D^{0}_{\rho \sigma}(k)=\frac{1}{k^2}\Big[\delta_{\rho\sigma}-\frac{k_\rho k_\sigma}{k^2} \Big]$ is the Landau-gauge free gauge-boson propagator, ${\cal G}(k^2)$ is a model effective interaction, and $\Gamma_{\sigma}(q,p)$ is the effective quark-gluon vertex. 
Since the chemical potential only appears explicitly in the DSE of quark propagator, its effects on the gluon propagator and quark-gluon vertex are indirect. 
In this work, we investigate the rainbow approximation for the vertex:
\begin{equation}
\label{vertex} \Gamma^{RB}_\sigma(q,p;\mu) = \gamma_{\sigma}^{} \, .
\end{equation}

For the model effective interaction in Eq.(\ref{KernelAnsatz}), Qin-Chang (QC) model~\cite{Alkofer:2002bp,QC2011} and Maris-Tandy(MT) model~\cite{Maris1999} have been widely employed in the context of DSE. The main difference between the two models lies in the deep infrared domain: the MT model vanishes when the gluon momentum goes to zero while the QC one to a finite constant. In this work,  we take the 'Infrared-QC' model which only express the long-range behavior of the renormalization group improved  Qin-Chang(QC) model, account of Lattice QCD simulations and gauge field theory analysis preferring the QC model\cite{Oliveira:2012eh,Aguilar:2012rz,Zwanziger:2012xg}.
Though the ultraviolet parts of the models ensure the correct perturbative behavior,
they are not essential in describing nonperturbative physics, e.g. spectrum of light mesons \cite{Alkofer:2002bp,Eichmann2016}.
Herein we neglect them since our work is based on the competition between the chiral condensate and quark number density in the infrared region \cite{Chen2009}.
The 'Infrared-QC' model is expressed as:
\begin{equation}
\label{IRGsQC} 
{\cal G}(k^{2}) = \frac{8\pi^2}{\omega^4} \, D\, {\rm e}^{-k^{2}/\omega^2} \, .
\end{equation}
Equations~(\ref{IRGsQC}) delivers an ultraviolet-finite model gap equation.
Hence, the regularization mass scale can be removed to infinity and the renormalization constants can be set to 1.
For the corresponding ans\"{a}tz at finite chemical potential, 
we follow that in Ref.~\cite{Chen2008}, neglecting the dependence of the effective interaction ${\cal G}$ and the gluon propagator on the chemical potential at low densities.

\section{The nonlocal quark condensate component of gluon vacuum polarization}
\label{SecIII}
\begin{figure}[htp!]
\vspace{-5mm}
\centering \includegraphics[width=0.35\textwidth]{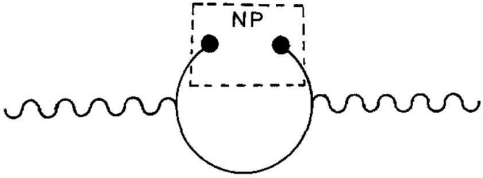}
\vspace*{-2mm}
\caption{Oder $g^2$ contribution to the gluon vacuum polarization the the quark-quark vacuum expectation value. The "NP" presents the nonlocal and nonperturbative non-local vacuum expectation of the quark field.}
\label{Fig_GluonSelfEnergy}
\end{figure}
The lowest-order, nonlocal quark scalar condensate component of the gluon vacuum polarization is an important field-theoretic test of OPE techniques, since the gluon vacuum polarization must be transverse in order to satisfy the Slavnov-Taylor identity. The OPE for the gluon propagator is\cite{T1989Quark}
\begin{eqnarray}
i D_{\mu \nu}^{ab}(p)  &=&  \int d^{4}xe^{ip \cdot x}
\langle 0|T(A_{\mu}^{a}(x)A_{\nu}^{b}(0))|0\rangle\nonumber \\
&=& \delta^{ab} C_{\mu \nu}^{I}(p) +C_{\mu \nu}^{\bar{\psi}\psi}(p)\delta^{ab}\langle \bar{\psi} \psi \rangle \nonumber \\
&& + C_{\mu \nu}^{FF}\delta^{ab}\langle F_{\mu \nu}^{c}F_{\mu \nu}^{c} \rangle \nonumber \\
&& + \ higher \ dimensional \ condensate
 \label{OPE1}
\end{eqnarray}

The first term in Eq.(\ref{OPE1}) represents the contribution of perturbation theory and the second term is the nonperturbative contribution of the interaction. The coefficient $C_{\mu \nu}^{\bar{\psi}\psi}(p)$ in this term can be evaluated perturbatively, as the OPE factors the long-distance behavior of QCD into condensates are of mass dimension less than ten\cite{Shifman:1978bx,Reinders:1984sr}. The lowest-order contribution to $C_{\mu \nu}^{\bar{\psi}\psi}(p)$ is represented by the vacuum polarization diagram in Fig.(\ref{Fig_GluonSelfEnergy}) and the amplitude\cite{T1989Quark} of Fig.(\ref{Fig_GluonSelfEnergy}) is:
\begin{eqnarray}
i D_{\alpha \beta}^{ab}(p)  &=&  -\frac{1}{2} g^{2} \int d^{4}x d^{4}y d^{4}z e^{ip \cdot x} \langle T(A_{\alpha}^{a}(x)A_{\nu}^{e}(y))\rangle\nonumber \\
&& \cdot Tr(\lambda ^{c} \gamma ^{\mu} \langle T(\psi(y) \bar{\psi}(z)) \rangle 
\gamma ^{\nu} \lambda ^{e} ) \langle (\bar{\psi}(y) \psi(z)) \rangle\nonumber \\
&& \cdot \langle T(A_{\nu}^{e}(z)A_{\beta}^{b}(0))\rangle
 \label{OPE2}
\end{eqnarray}
Where the sum represents the sum of all the quark flavors. It has also been known that the DSEs of QCD provide a natural approach to investigate the chiral symmetry restoration in vacuum  (see, e.g., Refs.~\cite{Roberts1990,Chang2007,Wang2013,QCDPT-DSE21-2}), the nonperturbative content in Eq.(\ref{OPE2}) could calculated from Dyson-Schwinger equation in the previous section:
\begin{equation}
\label{OPE3}
\langle T(\psi(y) \bar{\psi}(z)) \rangle =\langle T(\psi(p) \bar{\psi}(0)) \rangle=S(p) .
\end{equation}
\begin{align}
& \langle \bar{\psi}(y) \psi(z) \rangle  \nonumber \\
& =  \langle \bar{\psi}(w) \psi(0) \rangle \nonumber \\
& = 12 \int \frac{d^4 q}{(2 \pi)^{4}} \frac{B(\widetilde{q}^2)\times e^{iwq}}{\Vec{q}^{2}A^{2}(\widetilde{q}^2)+\widetilde{q}_{4}^{2}
C^{2} (\widetilde{q}^2) +B^{2}(\widetilde{q}^2)}
\label{OPE4}
\end{align}

With the result of the quark propagator and the nonlocal quark condensate from the solution of DSEs, the nonlocal quark condensate component of the gluon vacuum polarization is identified:
\begin{align}
& \Pi_{\mu \nu}^{ab}(p)=\delta ^{ab}  \Pi_{\mu \nu}(p) \\
& \Pi_{\mu \nu}(p)=(\frac{p^{\mu }p^{\nu}}{p^2}-g^{\mu \nu})\Pi(p) \\
& \Pi_{\mu \nu}^{ab}(p)=-g^{2} \delta ^{ab}  \int d^{4} w
e^{ip \cdot w} \int \frac{d^{4}q }{(2 \pi)^{4}} e^{-iq \cdot w} \langle \bar{\psi}(w) \psi(0) \rangle\nonumber \\
& \cdot Tr \Bigg \{ \gamma^{\mu} \frac{1}{i\vec{\gamma}\cdot
\vec{p}A(\widetilde{q}^2)+i\gamma_4\tilde{q_4}C(\widetilde{q}^2)
+ B(\widetilde{q}^2)} \gamma ^{\nu}  \Bigg \} 
\label{OPE5}
\end{align}

Finally, we can obtain the quark's polarization scalar with the lowest-order OPE model:
\begin{eqnarray}
 \Pi(p)&=&64 g^{2} \int d^{4} w \int \frac{d^{4}q }{(2 \pi)^{4}} \int \frac{d^{4}n}{(2 \pi)^4} \Bigg \{ \nonumber \\
&& e^{ip \cdot w} \times e^{-iq \cdot w} \times e^{iwn} \frac{B(\widetilde{q}^2)}{\Vec{q}^{2}A^{2}(\widetilde{q}^2)+\widetilde{q}_{4}^{2}
C^{2} (\widetilde{q}^2) +B^{2}(\widetilde{q}^2)}  \nonumber \\
&& \frac{B(\widetilde{n}^2)}{\Vec{n}^{2}A^{2}(\widetilde{n}^2)+\widetilde{n}_{4}^{2}
C^{2} (\widetilde{n}^2) +B^{2}(\widetilde{n}^2)} \Bigg \}
\label{OPE-quark-polarization}
\end{eqnarray}

The nonlocal quark condensate component of gluon vacuum polarization in Eq.(\ref{OPE-quark-polarization}) is more complicated but accurate than the local quark condensate component of gluon vacuum polarization in Eq.(\ref{old_OPE}), since the Eq.(\ref{OPE-quark-polarization}) doesn't depend on the quark current mass and can be used in the infrared domain. Compared with the previous work, the integral of Eq.(\ref{OPE-quark-polarization}) converges due to the behaviour of function $B(p^2)$ decreasing quickly to zero in the ultraviolet domain. It is worth noting that, the twelve integral in Eq.(\ref{OPE-quark-polarization}) makes it difficult to solve numerically. Compared with solving the Eq.(\ref{OPE-quark-polarization}) directly, we separate the integral of $\int d^{4} n$ from the twelve integral, the integral of $\int d^{4} n$ presents the result of the nonlocal quark condensate and should be calculated in advance.

In Landau gauge, the vacuum polarization modifies the gluon propagator in the following way:
\begin{equation}
\label{OPEGluonPropagator}
D^{\mu \nu }(p)=(\delta^{\mu \nu}-\frac{p^{\mu}p^{\nu}}{p^2})D(p)
\end{equation}

\begin{figure}[htp!]
	\centering
\subfigure
{
	\vspace{-0.2cm}
	\begin{minipage}{8cm}
	\centering
	\centerline{\includegraphics[scale=0.699,angle=0]{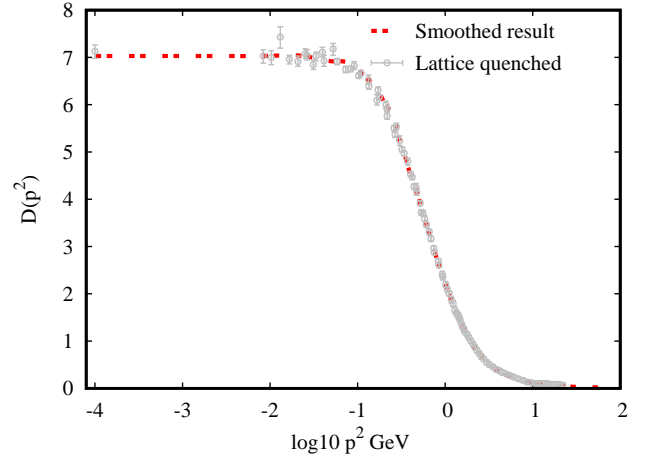}}
	\end{minipage}
}
\vspace{-0.2cm}

\subfigure
{
	\begin{minipage}{9cm}
	\centering
	\centerline{\includegraphics[scale=0.699,angle=0]{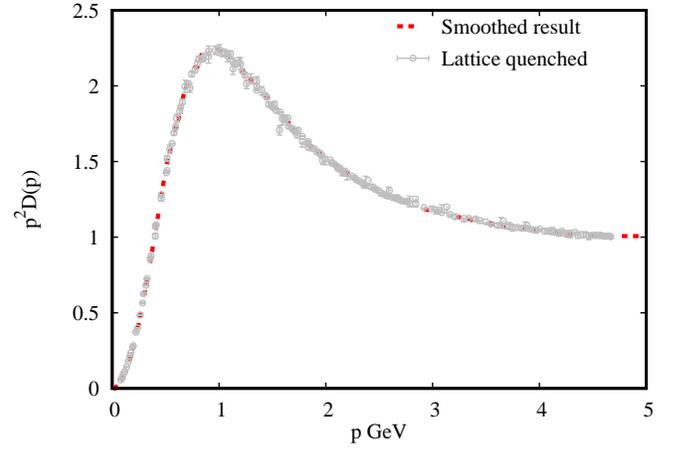}}
	\end{minipage}
}
\vspace{-0.2cm}
\vspace*{0mm} \caption{\label{lattice_smoothed} The lattice unquenched gluon propagator(upper) and dressing gluon propagator(under).}
\end{figure}

Our dressed gluon propagator is then given by the gluon vacuum polarization from OPE with quark' contribution and the pure("quenched") Yang-Mills propagator.
\begin{figure}[h]
\vspace*{0.0mm}
\centering \includegraphics[width=0.35\textwidth]{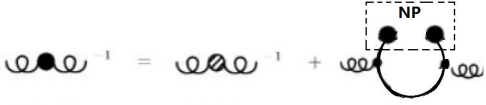}
\vspace*{0.0mm}
\caption{Quenching of gluon propagator with OPE approach. The shaded propagator is the quenched propagator from Lattice QCD.}
\label{Fig_GluonDSE}
\end{figure}
\begin{equation}
\label{OPEGluonPropagator1}
D(p)=\frac{1}{D_{YM}^{-1}(p^2)+\Pi_{q}(p^2)}
\end{equation}
where $D_{YM}(p^2)$ corresponds to the pure("quenched") Yang-Mills propagator, and we smoothed the lattice simulation data(shown in the Appendix) from article \cite{PhysRevD.86.074512} and the results are shown in Figure. \ref{lattice_smoothed}.

Implying that the finite value of the gluon propagator in the infrared domain  can be interpreted as setting a scale for the effective gluon mass $\mathcal{M}(p^2)=D(p \to 0)^{-1/2}$, it has been verified that the value of the dynamical gluon mass\cite{GayDucati:1993fn,Mihara:2000wf,2004A} in the infrared $\mathcal{M}(p^2=0)$
is frozen at the scale $m_g$= 1.2 - 2 $\Lambda_{QCD}$, which can be used to improve the comparison between experiment and theory in the case $J / \psi $ of  decays\cite{Parisi:1980jy}.

Finally, we investigate possible positivity violations in the gluon propagator.
According to the Osterwalder-Schrader axiom of reflection positivity, a two-point correlation function $\Delta(x-y)$ of Euclidean field theory has to fulfill the condition\cite{Fischer:2003rp,Alkofer:2003jj}:
\begin{equation}
\label{Sxcheinger1}
\int d^{4}x d^{4}y \overline{f} (\vec{x},-x_{0}) \Delta(x-y) f(\vec{y},-y_{0}) \geqslant 0
\end{equation}
if a physical particle is described. Here $f(\vec{x},-x_{0})$ are complex valued test functions. A violation of
this condition signals the absence of the corresponding particle from the physical spectrum of the theory, i.e. the particle is confined. After a three dimensional transformation Fourier transformation and for the special case $\vec{p}=0$, the Osterwalder-Schrader condition for gluon propagator can be given in terms of the Schwinger function $\Delta (t) $ defined by :
\begin{eqnarray}
 \Delta (t)&=&\int d^{3} x \int \frac{d^{4}p}{(2 \pi)^4} e^{i(tp_{4}+\vec{x} \cdot \vec{p})}D(p^2)\nonumber \\
&=& \frac{1}{\pi} \int dp_{4} cos(tp_{4})D(p_{4}^{2}) \geqslant 0
\label{Sxcheinger2}
\end{eqnarray}

\section{Numerical calculations and Results}
\label{SecIV}
\begin{figure*}[htp!]
	\centering
\subfigure
{
	\vspace{-0.2cm}
	\begin{minipage}{8cm}
	\centering
	\centerline{\includegraphics[scale=0.699,angle=0]{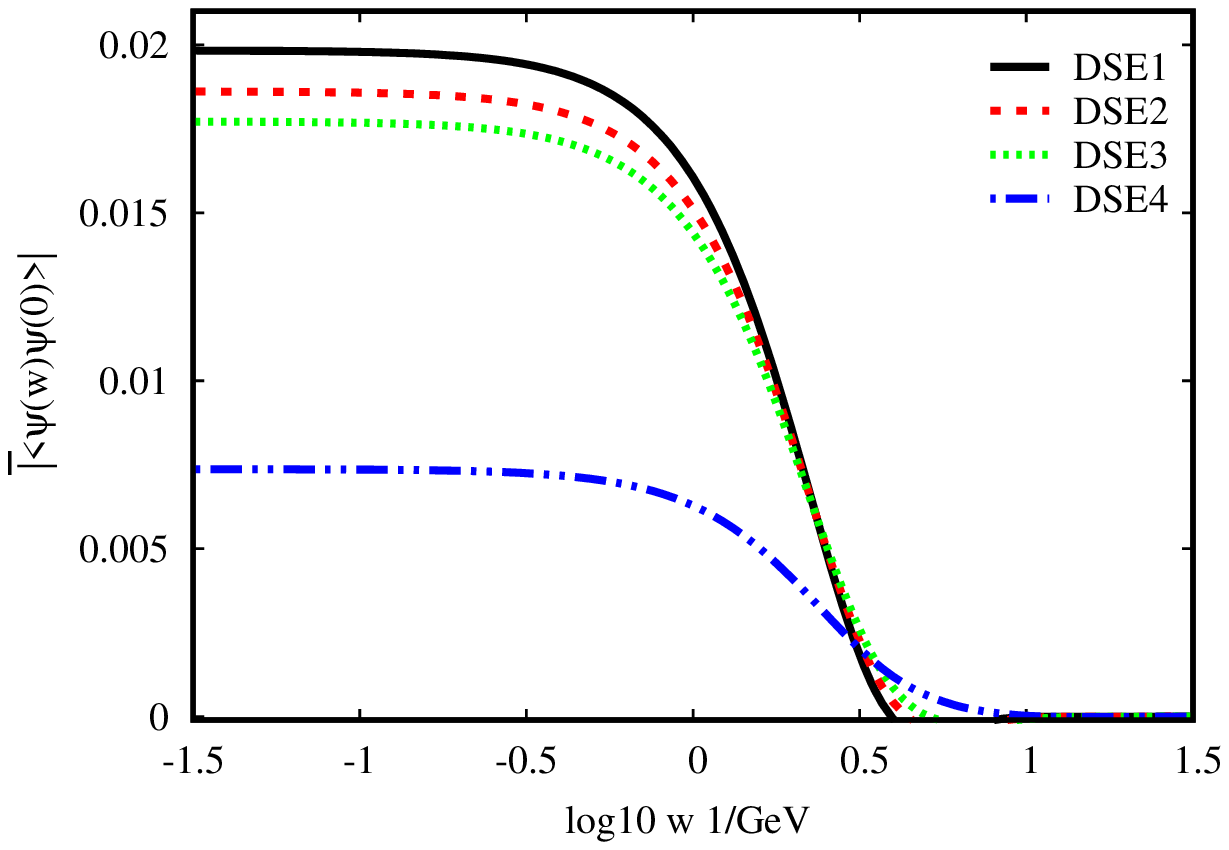}}
	\end{minipage}
}
\vspace{-0.2cm}
\subfigure
{
	\begin{minipage}{9cm}
	\centering
	\centerline{\includegraphics[scale=0.699,angle=0]{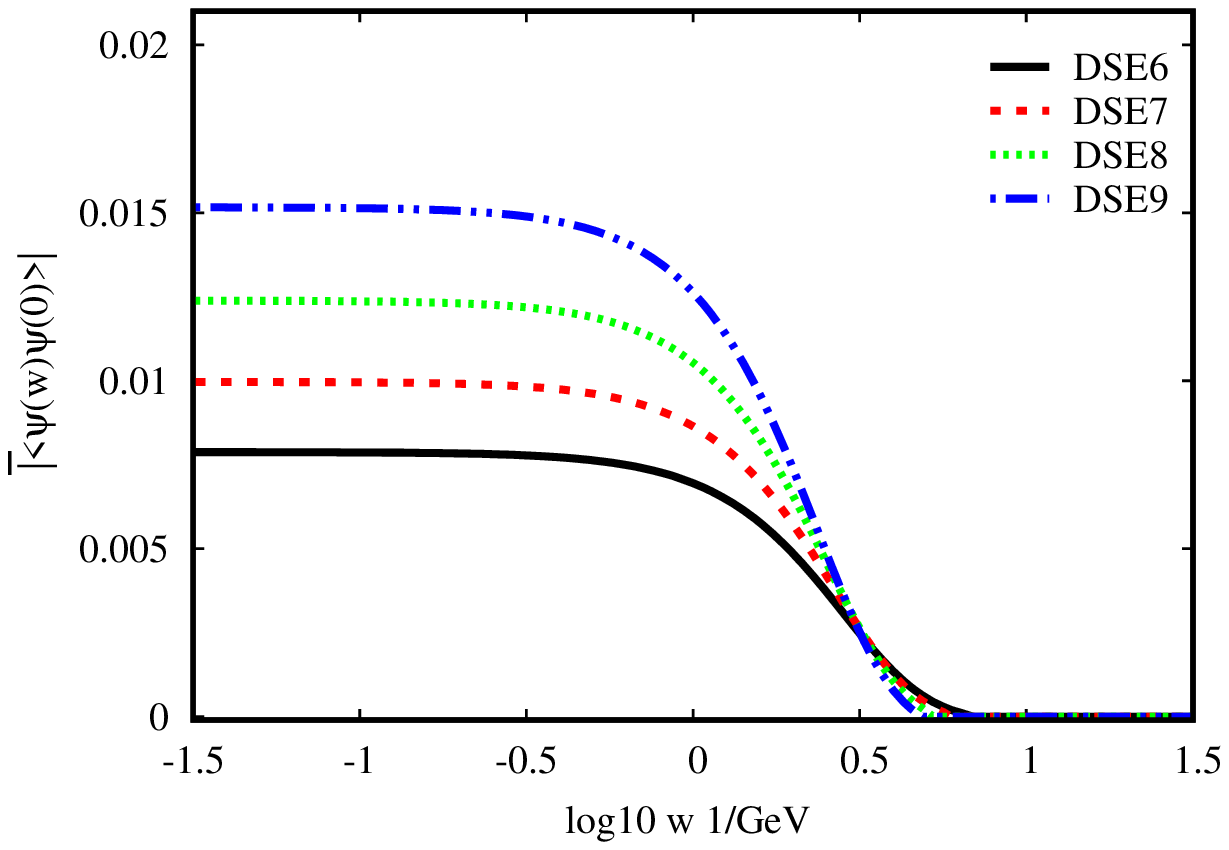}}
	\end{minipage}
}
\vspace{-0.2cm}
\vspace*{0mm} \caption{\label{Non-local_quarkcondensate} Non-local quark condensate with different model at vacuum.}
\end{figure*}

\begin{figure*}[htp!]
	\centering
\subfigure
{
	\vspace{-0.2cm}
	\begin{minipage}{8cm}
	\centering
	\centerline{\includegraphics[scale=0.699,angle=0]{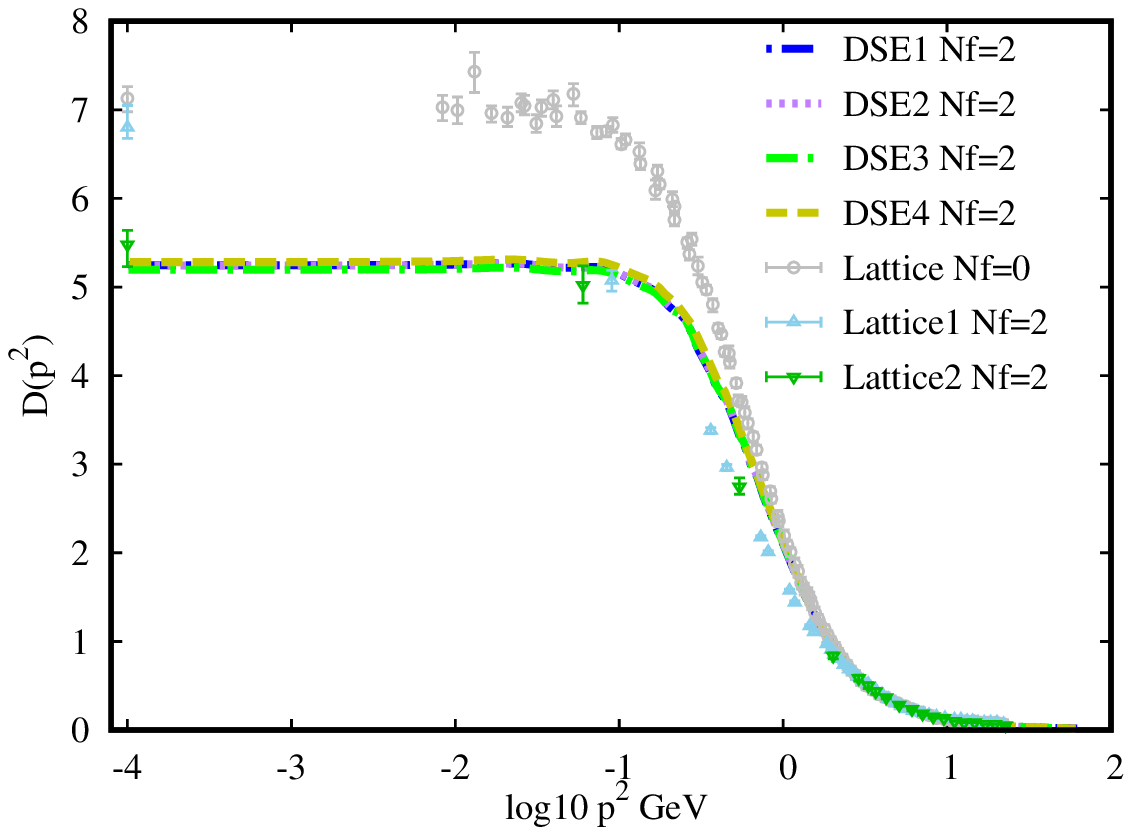}}
	\end{minipage}
}
\vspace{-0.2cm}
\subfigure
{
	\begin{minipage}{9cm}
	\centering
	\centerline{\includegraphics[scale=0.699,angle=0]{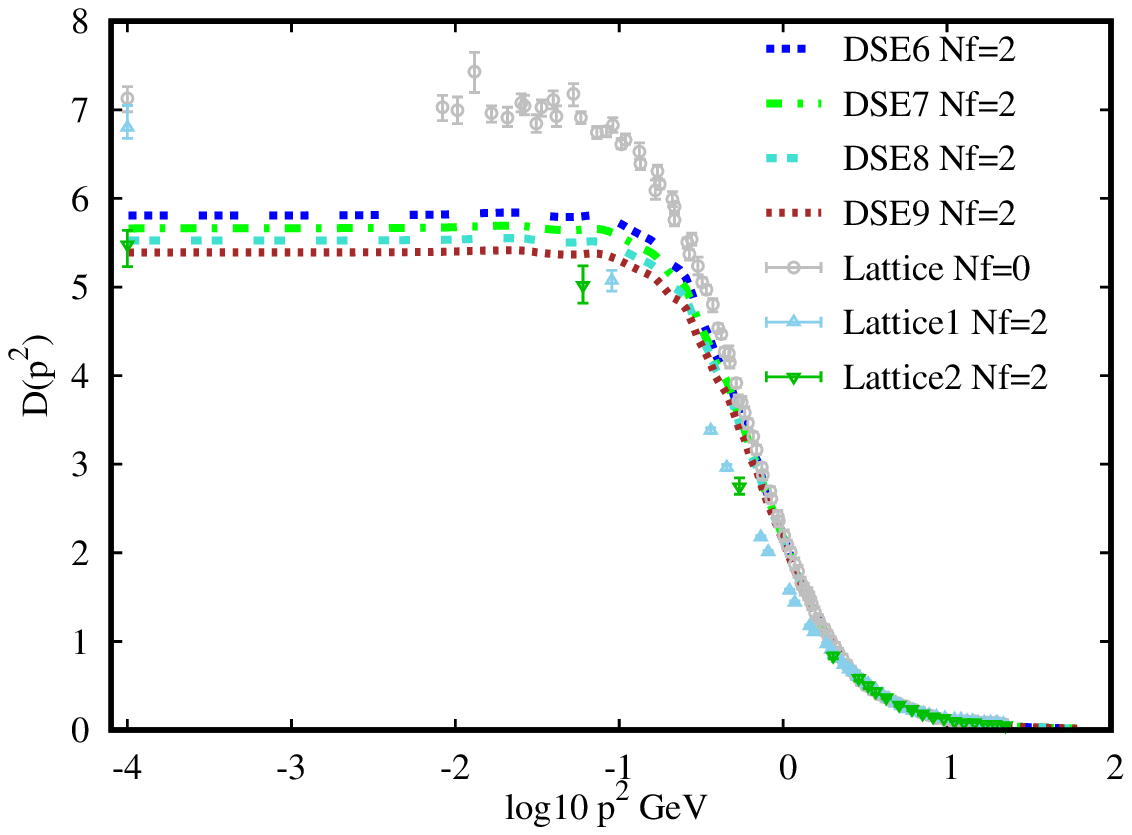}}
	\end{minipage}
}
\vspace{-0.2cm}
\subfigure
{
	\vspace{-0.2cm}
	\begin{minipage}{8cm}
	\centering
	\centerline{\includegraphics[scale=0.699,angle=0]{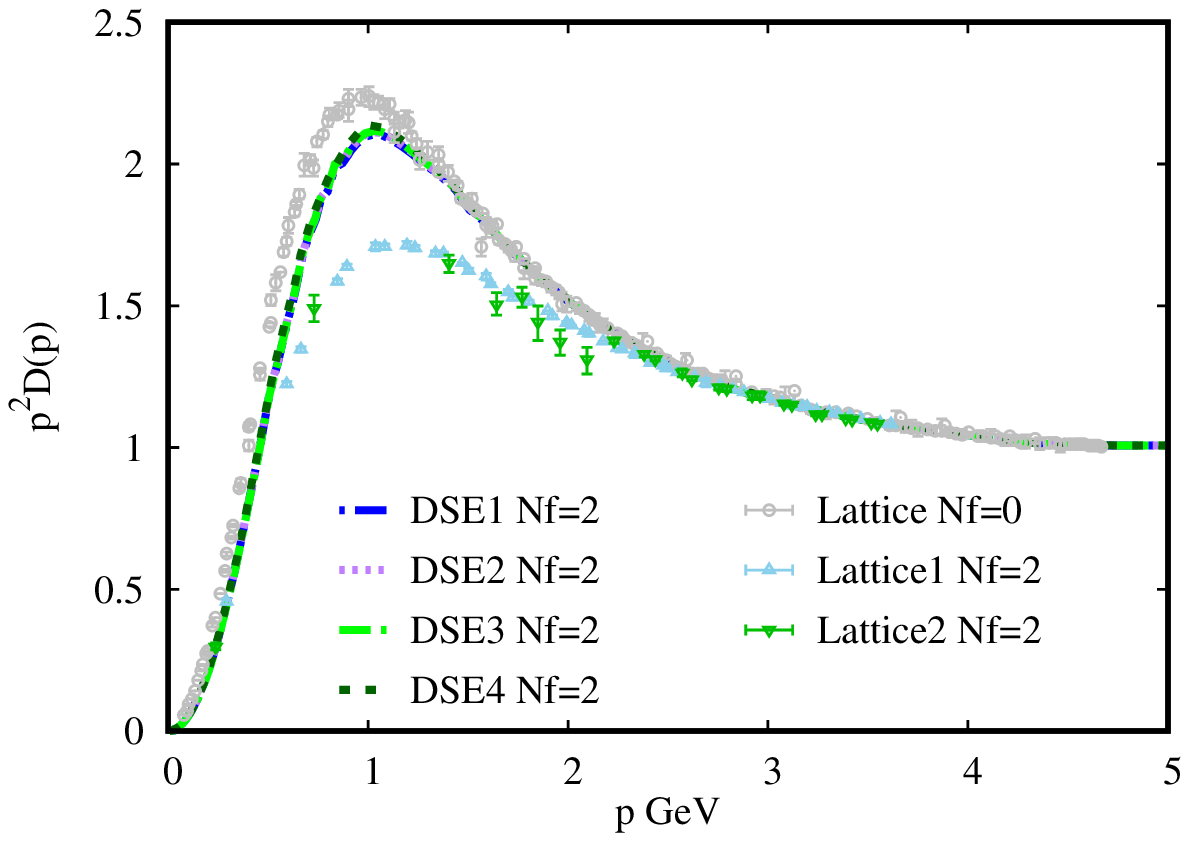}}
	\end{minipage}
}
\vspace{-0.2cm}
\subfigure
{
	\begin{minipage}{9cm}
	\centering
	\centerline{\includegraphics[scale=0.699,angle=0]{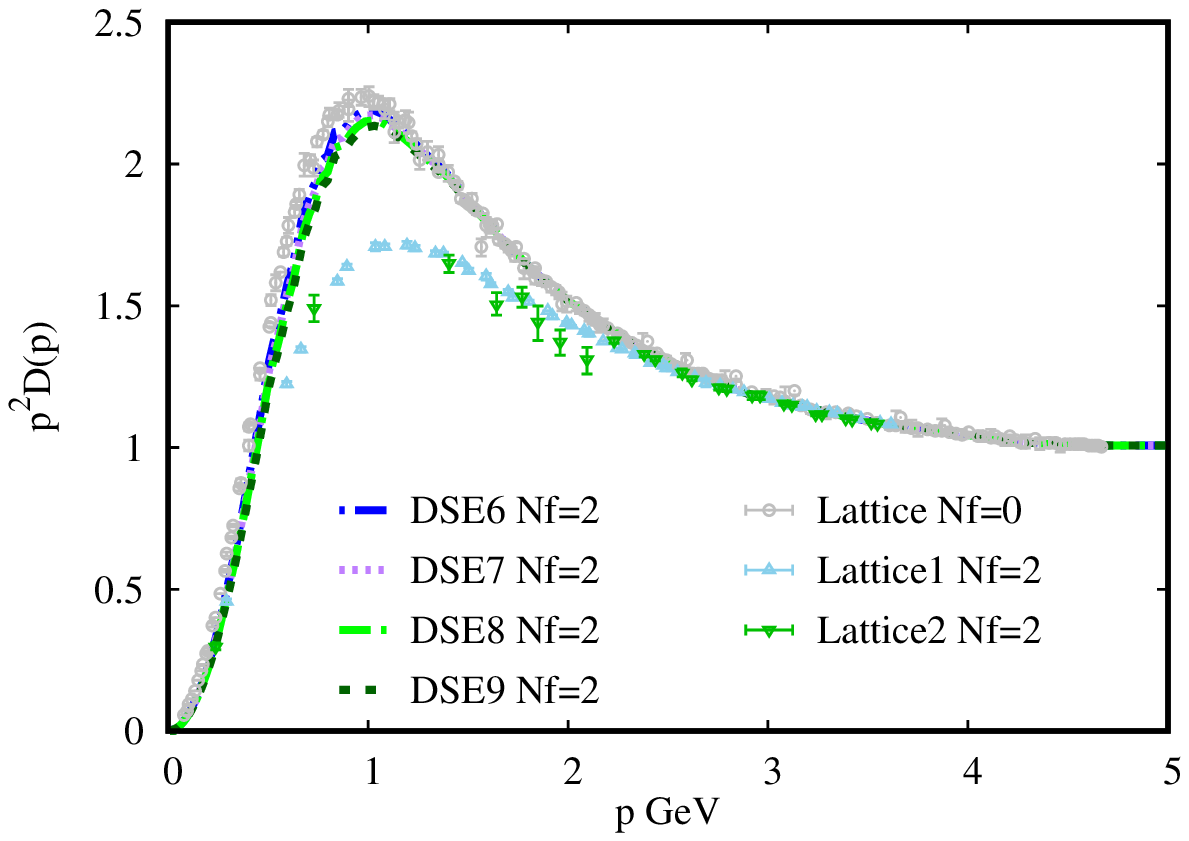}}
	\end{minipage}
}

\vspace*{0mm} \caption{\label{FiG_OPE-gluon-propagater}{The propagators and the dressing gluon propagators from our numerical results and the lattice simulation. The Lattice Nf=0 result presents the quenched result while the Lattice Nf=2 result presents the unquenched result.}}
\end{figure*}

\begin{figure*}[htp!]
	\centering
\subfigure
{
	\vspace{-0.2cm}
	\begin{minipage}{8cm}
	\centering
	\centerline{\includegraphics[scale=0.699,angle=0]{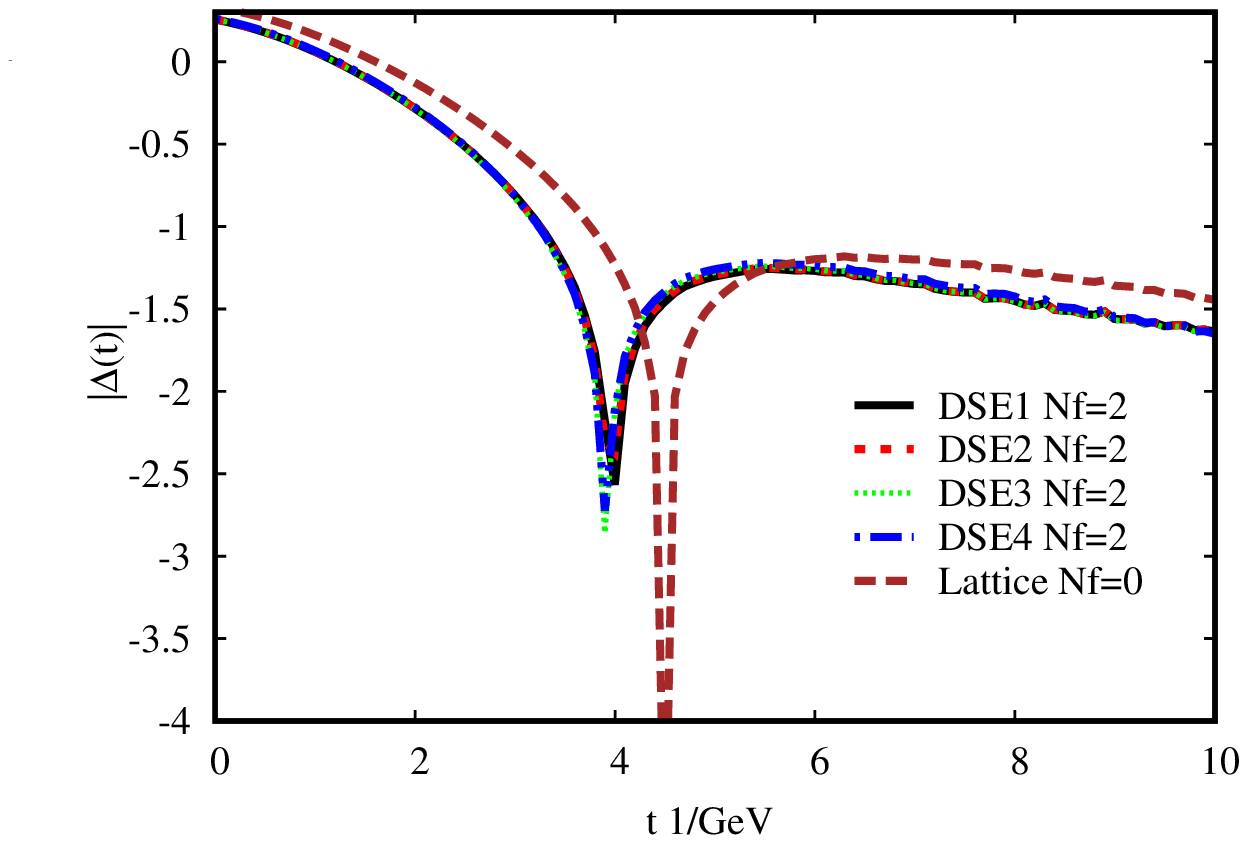}}
	\end{minipage}
}
\vspace{-0.2cm}
\subfigure
{
	\begin{minipage}{9cm}
	\centering
	\centerline{\includegraphics[scale=0.699,angle=0]{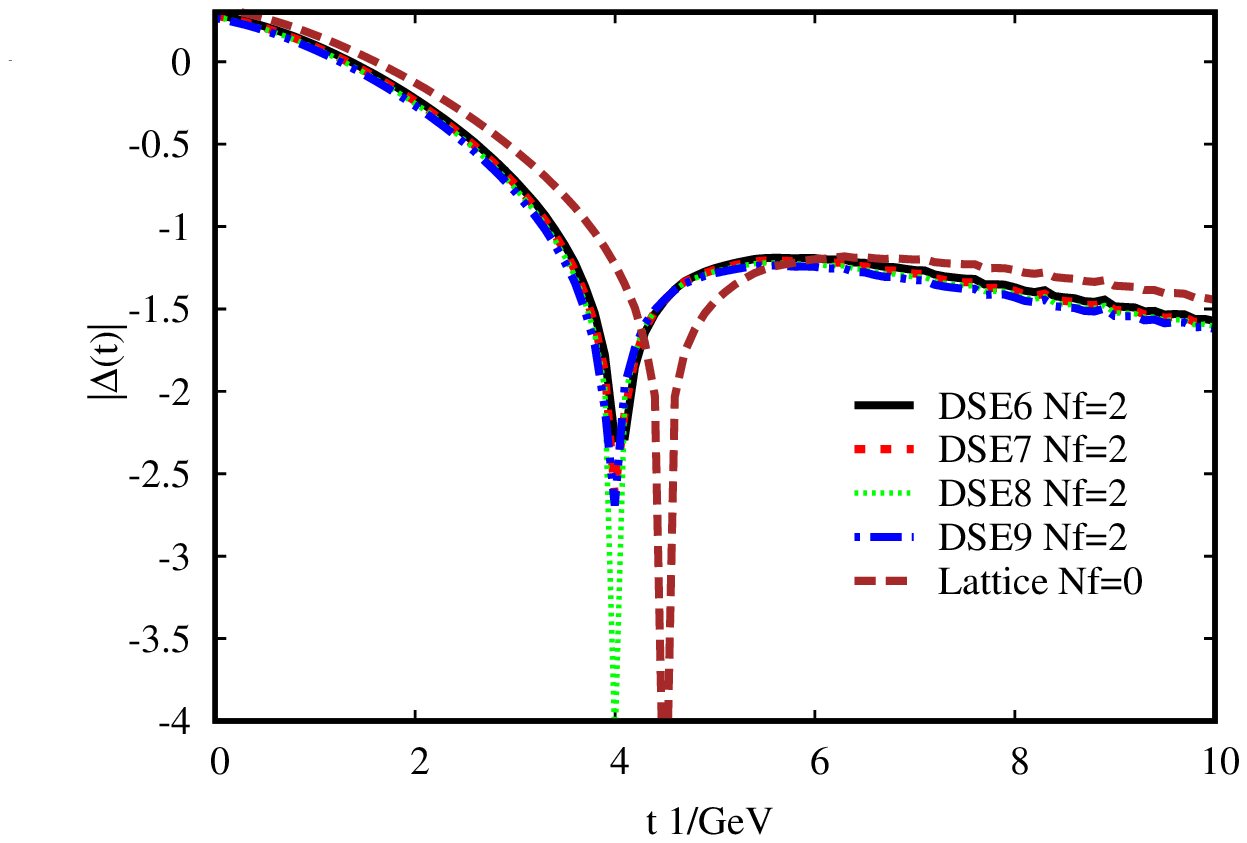}}
	\end{minipage}
}
\vspace{-0.2cm}
\vspace*{0mm} 
\caption{\label{Fig_SchwignerFUction}{The absolute value of the gluon Schwinger function from our calculations as well as from the lattice simulation\cite{Bowman:2007du}.}}
\end{figure*}

\begin{figure*}[htp!]
	\centering
\subfigure
{
	\vspace{-0.2cm}
	\begin{minipage}{8cm}
	\centering
	\centerline{\includegraphics[scale=0.699,angle=0]{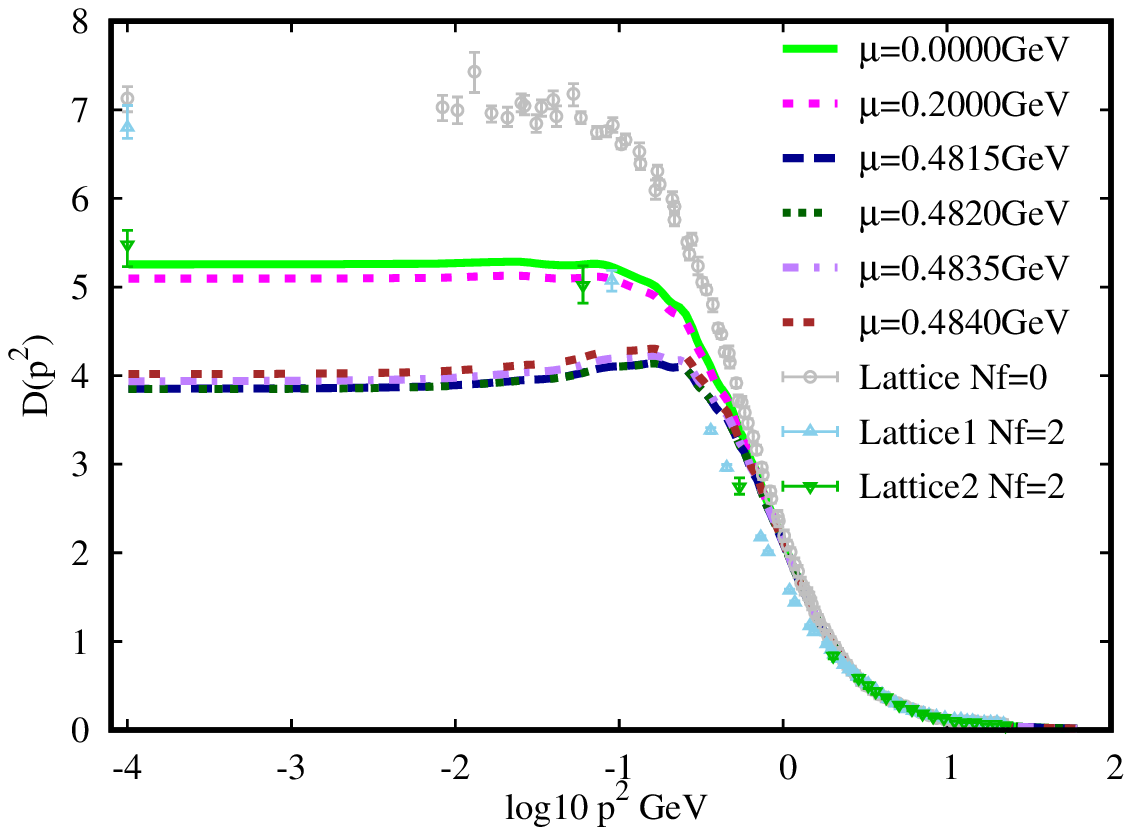}}
	\end{minipage}
}
\vspace{-0.2cm}
\subfigure
{
	\begin{minipage}{9cm}
	\centering
	\centerline{\includegraphics[scale=0.699,angle=0]{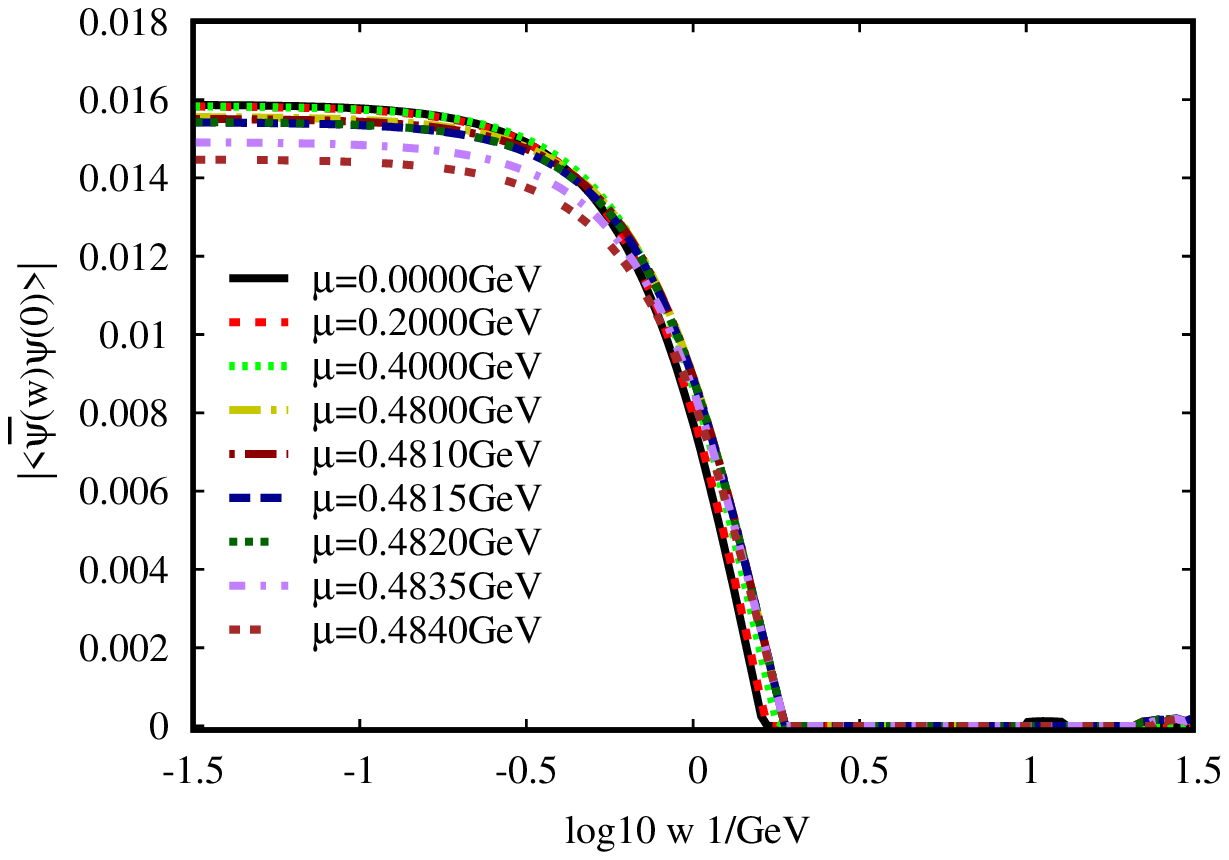}}
	\end{minipage}
}
\vspace{-0.2cm}
\subfigure
{
	\vspace{-0.2cm}
	\begin{minipage}{8cm}
	\centering
	\centerline{\includegraphics[scale=0.699,angle=0]{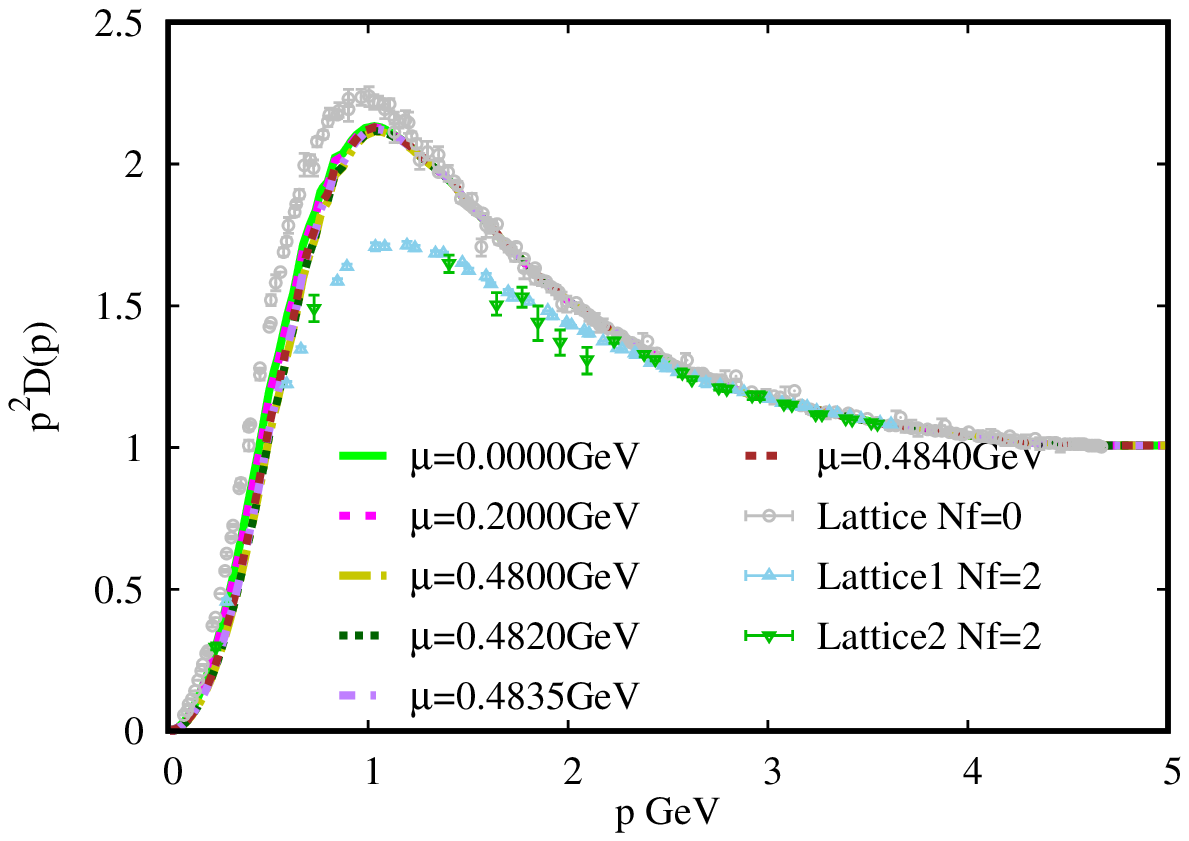}}
	\end{minipage}
}
\vspace{-0.2cm}
\subfigure
{
	\begin{minipage}{9cm}
	\centering
	\centerline{\includegraphics[scale=0.699,angle=0]{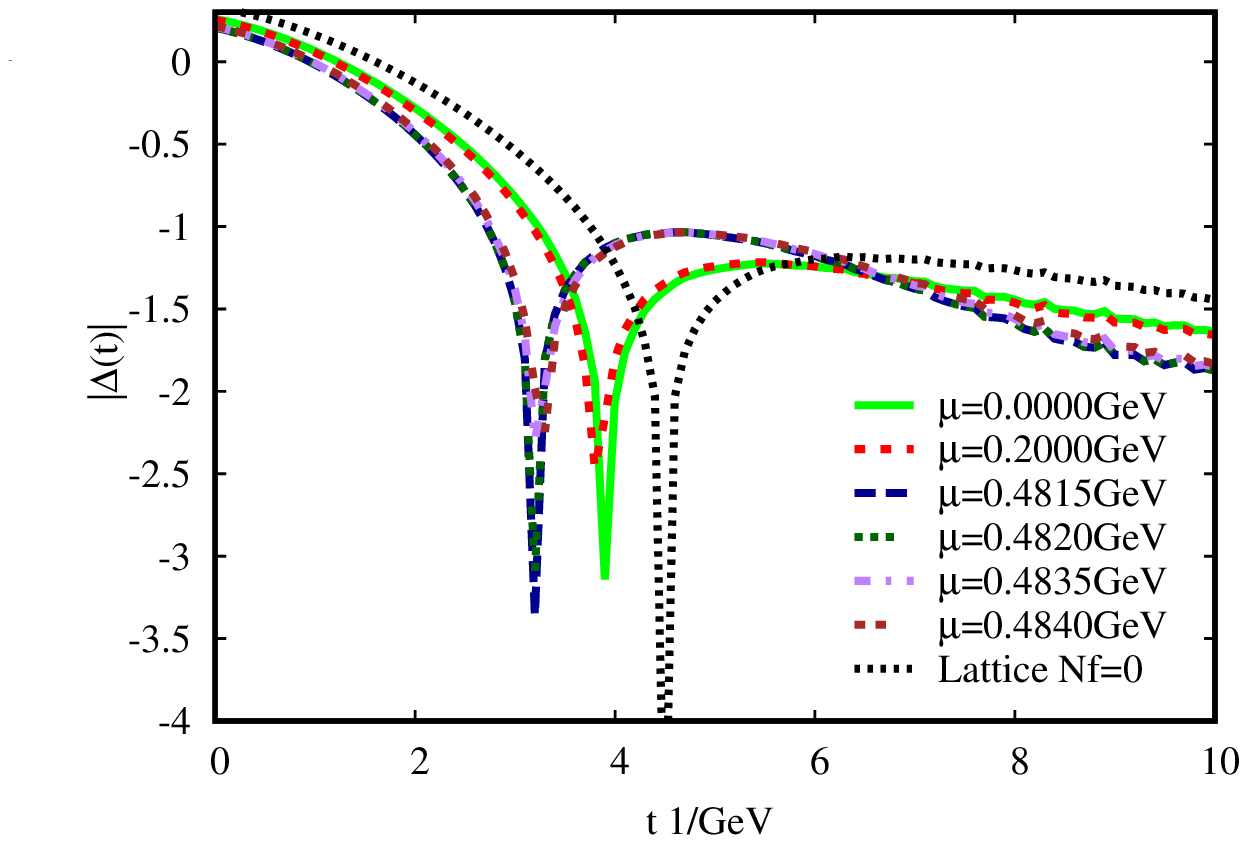}}
	\end{minipage}
}

\vspace*{0mm} \caption{\label{FiG_DSEchemical}{The gluon  propagators, non-local quark condensate, the dressing gluon propagators and the Schwinger functions from our numerical results with DSE5 at finite chemical potential and the corresponding result from lattice simulation.}}
\end{figure*}

\begin{table}[htp!]
\centering
\caption{Parameters and some characteristic numerical
results (dimensional quantities in units of GeV). DES1, DSE2, DSE3 and DSE4 are in the chiral limit $ m_q $ = 0, see the text for details. $-\langle \bar{\psi} \psi \rangle^{1/3}=-\langle \bar{\psi}(0) \psi(0) \rangle^{1/3}$ is the local quark condensate at vacuum. The values for the ratio $m_{g}/\Lambda_{QCD}$ are shown in the last column with the QCD scale $\Lambda_{QCD}$=0.312 GeV.}
\label{tab:1}       
\begin{tabular}{c@{\hspace{0.35cm}}c@{\hspace{0.35cm}}c@{\hspace{0.35cm}}c@{\hspace{0.35cm}}c@{\hspace{0.31cm}}c@{\hspace{0.31cm}}c@{\hspace{0.31cm}}}
\toprule
\toprule
model & $(\omega d)^{1/3}$ &$\omega$ & $D$ & $-\langle \bar{\psi} \psi \rangle^{1/3}$ & $m_g$ & $m_{g}/\Lambda_{QCD}$\\
\noalign{\smallskip}\hline\noalign{\smallskip}
DSE1 & 0.90& 0.40 & 1.86 &  0.271 & 0.435 & 1.39\\
DSE2 & 0.90& 0.50 & 1.49 &  0.264 & 0.435 & 1.39\\
DSE3 & 0.90& 0.60 & 1.24 &  0.260 & 0.435 & 1.39\\
DSE4 & 0.90& 0.70 & 1.06 &  0.223 & 0.435 & 1.39\\
DSE5 & 0.90& 0.67 & 1.10 &  0.250 & 0.436 & 1.39\\
DSE6 & 0.70& 0.50 & 0.69 &  0.200 & 0.415 & 1.33\\
DSE7 & 0.75& 0.50 & 0.84 &  0.214 & 0.420 & 1.34\\
DSE8 & 0.80& 0.50 & 1.02 &  0.228 & 0.425 & 1.36\\
DSE9 & 0.85& 0.50 & 1.23 &  0.246 & 0.430 & 1.38\\
\toprule
\toprule
\end{tabular}
\vspace*{-0.1cm}  
\end{table}

There are only two main parameters in our model:$D$ and $\omega$. We choose the set of values that can fit meson properties in vacuum well~\cite{Chang2009}. Calculations show that these observables are insensitive to the variation of $\omega \in $ [0.4, 0.7] GeV as long as $\omega D $ = constant. In our work, we chose $(\omega D)^{1/3} $ = 0.9 for DSE1, DSE2, DSE3 and DSE4 from  the article\cite{Qin:2011dd} and analyze how the parameter $\omega $ influence our result. In particular, the parameters $\omega \  $=0.678 GeV and $ D $ = 1.100 of DSE5 model is taken from the work\cite{Huang:2019crt}, which are fixed by calculating the quark chair condensate and the pion decay constant. In addition, we fixed the parameter $\omega $ in DSE6, DSE7, DSE8 and DSE9, then study how the parameter $(\omega D)^{1/3} $ change our results. It is worth nothing that, by following the previous work\cite{Huang:2019crt}, our work can be extended to calculate  $\sigma_s$ and further improve the results for $\sigma_{\pi N}^{}$.

With the above determined parameters and the evolution described in the last section,  we solve the Dyson-Schwinger equation of the quark propagator and calculate the nonlocal quark condensate. The obtained gluon propagator and the Schwinger function in the chiral limit are illustrated in the following.

In the Fig.(\ref{Non-local_quarkcondensate}), the left panel shows the nonlocal quark condensate with DSE1, DSE2, DSE3 and DSE4 in Table.(\ref{tab:1}) at vacuum, while the right panel presents the nonlocal quark condensate with DSE6, DSE7, DSE8 and DSE9. Our results show evidently that the condensate decreases gradually to zero with the increase of the space-time distance among the quarks definitely. We found the condensate is sensitive to the parameter $D$ and the condensate increases with the increase of $D$. 

In the Fig.(\ref{FiG_OPE-gluon-propagater}) we present the gluon propagators obtained with DSE1, DSE2, DSE3 and DSE4(left panel) and with DSE6, DSE7, DSE8 and DSE9 (right panel), and compare with the corresponding Lattice data. The behaviour of our obtained gluon propagators are alike with the Lattice results. Considering the approximate model in our work to calculate the gluon vacuum polarization, our obtained Nf=2 gluon propagators are located between the lattice Nf=0 gluon propagator and the lattice Nf=2 gluon propagator. The obtained gluon propagator of DSE1, DSE2, DSE3 and DSE4 from the left panel with the same value of $(\omega D)^{1/3} $ = 0.9 are almost unanimously, in spite of the huge difference of the non-local quark condensate in the left of Fig. \ref{Non-local_quarkcondensate}. On the other hand, by keeping the value of $(\omega ) $, we can investigate the effect of the parameter $\omega$ on the gluon propagator and and the $D(p=0, \mu=0)$ from Table.(\ref{tab:1}) . It is found that, with increasing $\omega$, the $D(p=0, \mu=0)$ gets bigger. In addition, we can obtain the ratio $m_{g}/\Lambda_{QCD}$ from the gluon propagator $D(p=0, \mu=0)$ are shown in the Table.(\ref{tab:1}) with the QCD scale $\Lambda_{QCD}$=0.312 GeV. The obtained results which range from 1.33 to 1.39 and are agreed with previous determinations for this ratio\cite{GayDucati:1993fn,Mihara:2000wf,2004A}. 
  
In the Fig.(\ref{Fig_SchwignerFUction}) we display the Fourier transform of the nontrivial part of the gluon propagator. The left panel are the results with DSE1, DSE2, DSE3 and DSE4 while  the right panel are the results with DSE6, DSE7, DSE8 and DSE9. The resulting positivity violation for the transverse gluon propagator in Landau gauge is a clear signal for gluon confinement. The zero crossing with different models occur at almost the same place in Euclidean time,t $\approx$ 1.14 fm =4.0 $GeV^{-1}$(with the relationship 1 $GeV^{-1}$  = 0.287 fm), which is about the size of a hadron. Due to the effect of quark loop, the zero crossing showed up early about 0.5 $GeV^{-1}$ = 0.14 fm compared with the lattice Nf=0 result. In other words, the size of a hadron deceases about 0.14 fm on account of considering the gluon vacuum polarization.
It's worth noting that for the Schwinger function $\Delta (t)$, our numerical results don't depend on the results of DSE due to the consistency and superiority of DSE approach.  

In the end, in Fig.(\ref{FiG_DSEchemical}), we expand the above conclusion to the case at finite chemical potential with parameters of DSE5, the change of $\langle \bar{\psi}(0) \psi(0) \rangle$, $\langle \bar{\psi}(p) \psi(0) \rangle$, the gluon propagators and the Schwinger functions dependence of the chemical potential are consist with our previous work\cite{Huang:2019crt,Chen2008}. More specifically, when the quark chemical potential $\mu < M_1$, where $M_1 = $ 0.478 GeV is the first constituent-mass-like pole of the quark propagator, the chiral condensate $\langle \bar{\psi}(0)  \psi(0) \rangle$ and $\langle \bar{\psi}(p) \psi(0) \rangle$ keep the same value as that in vacuum (i.e., at $\mu=0$), i.e. the system remains the same as that in the vacuum and no matter emerges\cite{Chen2008}. When $\mu > M_1$, the condensate decreases gradually. This indicates that dynamical chiral symmetry is partially restored in the medium at low density~\cite{ChangChen2007,Chen2009}. It is worth noting that the chemical potential dependence of the gluon vacuum polarization and the dressed gluon propagator can be obtained in our scheme. The results show it possible to calculate the light quark loop in Fig.(\ref{Fig_GluonDSE}) by eliminating the appearance of quadratic divergencies.

\section{Summary and Remark}
\label{SecV}
In summary, by using the Dyson-Schwinger equations approach of QCD in the chair limit we calculated the lowest-order, non-local quark condensate. Then, the nonlocal quark condensate component of gluon vacuum polarization based on the operator-product expansion technique could be obtained. And on this basis the unquenched gluon propagator in the  Landau gauge, the single gluon mass and the gluon's Schwinger function are obtained. Our obtained Nf=2 gluon propagators are located between the lattice Nf=0 gluon propagator and the lattice Nf=2 gluon propagator. Our results of the ratio $m_{g}/\Lambda_{QCD}$ which ranges from 1.34 to 1.73 agree with previous determinations for this ratio. And our numerical analysis of the gluon' Schwinger function finds clear evidence of the positivity violations in the gluon propagator. 

In our calculations, a method for obtaining the chemical potential dependence of the gluon vacuum polarization and the dressed gluon propagator is developed. Nonetheless, this work is only appropriate for the quark and gluon at low chemical potential and zero temperature. Since the gluon propagator will be separated into two parts: $D_{T}(p)$ and $D_{L}(p)$, the more complex quark condensate $p^{\mu}\langle \bar{\psi}(p)\gamma^{\mu} \psi(0) \rangle$ needs to be considered.

In this work, we find that the nonlocal quark condensate component of gluon vacuum polarization is more complicated but accurate than the local quark condensate component of gluon vacuum polarization in Eq.(\ref{old_OPE}), since the Eq.(\ref{OPE-quark-polarization}) doesn't depend on the quark current mass and can be used in the infrared domain. The benefits of the nonlocal quark condensate component of gluon vacuum polarization makes it possible to calculate the light quark loop in Fig.(\ref{Fig_GluonDSE}) by eliminating the appearance of quadratic divergencies. In this case, by following the previous work\cite{Huang:2019crt} and considering the effect of quark loop on the quark's DSE,  extending our work to calculate  $\sigma_s$ and further improving the results for $\sigma_{\pi N}^{}$ are under progress.

%
\section{Authors contributions}
All the authors were involved in the preparation of the manuscript. All the authors have read and approved the final manuscript.

\section*{Acknowledgements}
This study is financially supported by the National Key R$\&$D Program of China (No. 2018YFC1503705).
We acknowledge financial support from National Natural Science Foundation of China (Grants, No. G1323519204). The project was supported by the Fundamental Research Funds for National Universities,China University of Geosciences(Wuhan).

\section{Appendix}

\begin{table}[htp!]
\centering
\caption{The smoothed data of the Unquenched Lattice gluon propagator.}
\label{tab:2}       
\begin{tabular}{c@{\hspace{0.35cm}}c@{\hspace{0.35cm}}c@{\hspace{0.35cm}}c@{\hspace{0.35cm}}}
\toprule
\toprule
$p^{2}$ & $p^{2}D(p^{2})$ &$p^{2}$ & $p^{2}D(p^{2})$ \\
\noalign{\smallskip}\hline\noalign{\smallskip}
1.00000E-02 &7.02929E-04 & 1.77828E-02 & 2.22286E-03 \\
1.77828E-02 &2.22286E-03 & 1.98098E+00 & 1.53128E+00 \\
3.16228E-02 &7.02929E-03 & 2.03645E+00 & 1.49839E+00 \\
5.62341E-02 &2.22286E-02 & 2.09192E+00 & 1.47962E+00 \\
9.14195E-02 &5.87475E-02 & 2.16323E+00 & 1.44204E+00 \\
1.61359E-01 &1.83019E-01 & 2.24247E+00 & 1.40446E+00 \\
2.41611E-01 &4.03502E-01 & 2.31379E+00 & 1.37159E+00 \\
3.02357E-01 &6.22724E-01 & 2.37718E+00 & 1.34812E+00 \\
3.30735E-01 &7.30459E-01 & 2.44057E+00 & 1.32936E+00 \\
3.64492E-01 &8.67168E-01 & 2.51189E+00 & 1.30119E+00 \\
4.13888E-01 &1.08086E+00 & 2.59113E+00 & 1.27773E+00 \\
4.56132E-01 &1.24660E+00 & 2.66244E+00 & 1.25427E+00 \\
5.25750E-01 &1.53126E+00 & 2.74168E+00 & 1.23082E+00 \\
5.49870E-01 &1.58389E+00 & 2.84469E+00 & 1.20268E+00 \\
5.83759E-01 &1.69389E+00 & 2.93185E+00 & 1.18394E+00 \\
6.33794E-01 &1.81519E+00 & 3.01109E+00 & 1.16989E+00 \\
6.82992E-01 &1.98302E+00 & 3.09033E+00 & 1.16056E+00 \\
7.19686E-01 &2.02845E+00 & 3.16957E+00 & 1.14651E+00 \\
7.47096E-01 &2.07381E+00 & 3.24089E+00 & 1.13246E+00 \\
7.69775E-01 &2.12228E+00 & 3.32805E+00 & 1.11843E+00 \\
7.81371E-01 &2.11518E+00 & 3.42314E+00 & 1.10440E+00 \\
8.08240E-01 &2.15605E+00 & 3.49445E+00 & 1.09505E+00 \\
8.55780E-01 &2.18904E+00 & 3.56577E+00 & 1.08571E+00 \\
9.11250E-01 &2.21733E+00 & 3.64501E+00 & 1.07637E+00 \\
9.66720E-01 &2.23620E+00 & 3.71632E+00 & 1.07173E+00 \\
1.04596E+00 &2.22216E+00 & 3.78764E+00 & 1.06239E+00 \\
1.13312E+00 &2.16577E+00 & 3.85895E+00 & 1.05775E+00 \\ 
1.22029E+00 &2.09997E+00 & 3.93819E+00 & 1.04841E+00 \\
1.29160E+00 &2.04357E+00 & 4.00951E+00 & 1.04377E+00 \\
1.35499E+00 &2.00128E+00 & 4.08875E+00 & 1.03444E+00 \\
1.39461E+00 &1.97308E+00 & 4.16006E+00 & 1.02509E+00 \\
1.45008E+00 &1.92607E+00 & 4.23138E+00 & 1.02045E+00 \\
1.49762E+00 &1.86964E+00 & 4.29477E+00 & 1.01581E+00 \\
1.55309E+00 &1.83205E+00 & 4.35816E+00 & 1.01587E+00 \\
1.61648E+00 &1.78976E+00 & 4.42948E+00 & 1.01123E+00 \\
1.66403E+00 &1.73333E+00 & 4.50872E+00 & 1.00660E+00 \\
1.73534E+00 &1.69105E+00 & 4.58003E+00 & 1.00666E+00 \\
1.79081E+00 &1.64874E+00 & 4.65135E+00 & 1.00673E+00 \\
1.85420E+00 &1.60645E+00 & 4.58003E+00 & 1.00666E+00 \\
1.92552E+00 &1.56417E+00 & 4.65135E+00 & 1.00673E+00 \\
\toprule
\toprule
\end{tabular}
\vspace*{-0.1cm}  
\end{table}
%
%
\bibliographystyle{ws-mpla}
\bibliography{reference}

\end{document}